# Comment on "Reproducibility study of Monte Carlo simulations for nanoparticle dose enhancement and biological modeling of cell survival curves" by Velten et al. [Biomed Phys Eng Express 2023;9:045004]

Hans Rabus
Physikalisch-Technische Bundesanstalt, Berlin, Germany
**Email**: hans.rabus@ptb.de

**Abstract**

This comment highlights two methodological issues with the recent article by Velten et al. [Biomed Phys Eng Express 2023;9:045004]

**Introduction**

In their recent article, Velten et al. [1] replicated the physical simulation and biological modeling of previously published research [2] on 50 nm gold nanoparticles (GNPs) irradiated with different photon spectra or with protons in a spread-out Bragg peak (SOBP). For the biological modeling, they used a variant of the local effect model (LEM), which they called the "LEM approach in 2D" and which is based on the assumption of an infinitesimally thin cell. This comment is to point out that (a) this "LEM approach in 2D" leads to biased results, and (b) there is a conceptional problem with the quantity referred to as a "macroscopic dose".

**Concerns regarding the "LEM approach in 2D"**

The LEM in its original form [3] is based on the assumption that the survival rate $S$ of a cell line at an absorbed dose $D$ administered by low linear energy transfer (LET) irradiation is given by

$$S(D) = e^{-\langle N_p(D)\rangle} \quad (1)$$

where $\langle N_p(D)\rangle$ is the expected number of lethal lesions in the nucleus (at this dose and for this low-LET radiation quality). For low-LET radiation, the statistical expectation of the energy imparted in a small subvolume of the nucleus is assumed to be uniformly distributed across the nucleus. Similarly, the probability that a lesion will occur in such a small subvolume is also assumed to be uniformly distributed over the nucleus.

Therefore, if the irradiation is such that the expected value of the energy imparted varies across the dimensions of the cell nucleus, as for ion beams, the expected value of the number of lethal lesions is given by Eq. (2).

$$\langle N(D;\alpha,\beta)\rangle = \iiint_{V_n} \frac{N_p(D(x,y,z);\alpha,\beta)}{V_n} dV \quad (2)$$

Here, $V_n$ is the volume of the nucleus and $\alpha,\beta$ are the parameters of the linear-quadratic (LQ) model describing cell survival as a function of dose for uniform irradiation with low-LET radiation. $\langle N(D;\alpha,\beta)\rangle$ is thus the integral of the spatial density of the expected lesions.

In the "LEM approach in 2D" of Velten et al., an infinitely thin cell is assumed and Eq. (2) is replaced by

$$\langle N(D;\alpha,\beta)\rangle = \iint_{A_n} \frac{N_p(D(x,y);\alpha,\beta)}{A_n} dA \quad (3)$$

where $A_n$ is the cross-sectional area of the nucleus.

Although not explicitly stated, the description in the Materials and Methods section suggests that the further steps of their procedure were as follows. First, a number of points equal to the expected number of nanoparticles undergoing a photon or proton interaction were randomly selected in the infinitesimally thin cell. Then the dose kernels obtained from their simulations were placed at these positions to obtain the two-dimensional dose distribution, which was then used in Eq. (3) to obtain the expected number of lethal lesions.

If this interpretation of the procedure is correct and if the dose kernels shown in Supplementary Fig. 5 of Velten et al. (2023) were used, this leads to an overestimation of the number of lesions and an underestimation of cell survival. This is because using a cross-section of the 3D dose profile is equivalent to assuming that the cell has the shape of a cylinder in which the dose profile is constant along the direction of the cylinder axis. (As is approximately true for ion beams with energies much higher than the Bragg peak energy.) But then, instead of a nanoparticle undergoing an interaction, which is approximately a point source, one would have a line source. This would result in a much higher average dose in the cell nucleus than would be the case for nanoparticles.

Using a cut through the 3D dose map instead of an Abel transform from 3D to 2D means that the dose profile has both wrong magnitude and wrong dependence on radial distance from the GNP. Transforming the 3D dose distribution to 2D is equivalent to averaging over the third dimension. With a $1/r^2$ dose dependence in 3D, the transformed distribution has a $1/r$ dependence, and the peak dose at the surface of the GNP is reduced by a factor proportional to the ratio of GNP diameter to the thickness of the nucleus. The proportionality constant is of the order of $\pi$. For a GNP radius of 25 nm and a nucleus diameter of 8 µm, the reduction factor is thus about 0.01, which means that the dose at the surface of the GNP is 3 Gy for the 83 keV photons (which produce the highest surface dose, cf. Table S2 in [1]) instead of 300 Gy.

Since nothing of the preceding is mentioned in the work of Velten et al., it must be assumed that a central cut through the 3D dose distribution was used and that the results are overestimated as explained in the following.

*An illustrative example: a single nanoparticle*

To illustrate the argument above, consider the case where the GNPs are inside the nucleus and only one GNP undergoes an interaction. Assume further that the extra dose due to an ionization in the GNP varies with the inverse square of the radial distance from the GNP center. Then the dose inside this cell nucleus is essentially equal to $D_0$ everywhere except inside a sphere of radius $r_m$ around the GNP, where the dose significantly exceeds $D_0$ and is given by Eq. (4).

$$D = D_0 + D_1 \frac{r_s^2}{r^2} \qquad (4)$$

Here $D_1$ is the excess dose at the surface of the spherical GNP with radius $r_s$, and the origin of the coordinate system is at the center of the GNP. Assume further that the GNP is located such that aforementioned sphere of radius $r_m$ is completely inside the cell nucleus.

For simplicity, we further assume that the dose at the surface of the GNP is below the threshold dose for the transition from the LQ dose dependence of cell survival to linear high-dose dependence. Then the expected number of lesions for the 3D case is obtained as

$$\langle N_3(D) \rangle = \alpha D_0 + \beta D_0^2 \\ + (\alpha + 2\beta D_0) D_1 \frac{4\pi r_s^2 (r_m - r_s)}{V_n} \\ + \beta D_1^2 \frac{4\pi r_s^3}{V_n} \frac{(r_m - r_s)}{r_m} \qquad (5)$$

For the 2D case, assume that the nucleus has a cross section $A_n$ and that the GNP is located such that the circle of radius $r_m$ around the GNP is completely inside the nucleus. Then the dose in the nucleus is $D_0$ outside the circle, while the dose inside the circle is given by Eq. (4), where the origin of the coordinate system is again at the center of the GNP. The expected number of lesions is then obtained as

$$\langle N_2(D) \rangle = \alpha D_0 + \beta D_0^2 \\ + (\alpha + 2\beta D_0) D_1 \frac{2\pi r_s^2}{A_n} \ln\left(\frac{r_m}{r_s}\right) \\ + \beta D_1^2 \frac{\pi r_s^2}{A_n} \frac{(r_m^2 - r_s^2)}{r_m^2} \qquad (6)$$

The ratio $R_3$ between the third summand in Eq. (6) and the third summand in Eq. (5) is

$$R_3 = \frac{h_n}{2(r_m - r_s)} \ln\left(\frac{r_m}{r_s}\right) \qquad (7)$$

where $h_n = V_n/A_n$, that is, the average thickness of the nucleus along the third dimension.

The corresponding ratio $R_4$ between the fourth summands in Eq. (6) and Eq. (5) is

$$R_4 = \frac{h_n}{4r_s} \frac{(r_m + r_s)}{r_m} \qquad (8)$$

In Fig. 1 of Velten et al., the dose around a GNP undergoing an interaction falls to below 0.5 Gy within 250 nm of the GNP center. For the photon irradiation, the decrease between the surface of the GNP and this distance is even more than two orders of magnitude, since the gold M-shell Auger electrons are completely stopped in this range. Thus, 250 nm appears to be a reasonable lower limit for $r_m$. If the sphere and the circle both have a diameter of 8 µm, as considered by Velten et al. [1], then $h_n$ is about 5.3 µm.

Using these values in Eqs. (7) and (8), $R_3$ and $R_4$ are obtained as approximately 30 and 60, respectively. This means that the number of lesions originating from the GNP is overestimated by orders of magnitude in this case when the "LEM approach in 2D" is used instead of the conventional LEM.

*Adaptation of the example to many nanoparticles*

The case of many GNPs instead of just one is more complex. Since the GNPs are assumed to be uniformly distributed in the cell nucleus, a modified version of Eq. (4) can be used for the dose in the vicinity of the GNP:

$$D = D_b + D_1 \frac{r_s^2}{r^2} \quad (9)$$

Here $D_b$ is the background dose, which also contains the contributions from the other GNPs undergoing an ionizing interaction. As is shown in Fig. S1.1 (Supplement 1), this background dose varies over the cell nucleus by up to a factor of 2 (in the 3D case). The mean values for the 3D and 2D cases (cf. Supplement 1) are given by Eqs. (10) and (11).

$$\bar{D}_{b,3} = D_0 + N_i D_1 \frac{r_s^2}{R_n^2} \times \left(\frac{9}{4} - \frac{3}{N_i^{1/3}}\right) \quad (10)$$

$$\bar{D}_{b,2} = D_0 + N_i D_1 \frac{r_s^2}{R_n^2} \times \ln(0.68 N_i) \quad (11)$$

Here, $D_0$ is the dose in the absence of GNPs, $R_n$ is the radius of the nucleus, and $N_i = NpD_0$ is the number of GNPs undergoing an ionizing interaction. $N$ is the total number of GNPs in the nucleus, and $p$ is the probability per dose of a GNP undergoing an ionizing interaction.

Neglecting the variation of $D_b$ and using only the mean value $\bar{D}_b$, the necessary adjustments to Eqs. (4) and (5) are to replace $D_0$ by $\bar{D}_b$, $D_1$ by $N_i D_1$, and $D_1^2$ by $N_i D_1^2$, which leads to Eqs. (12) and (13).

$$\langle N_3(D) \rangle = \alpha \bar{D}_{b,3} + \beta \bar{D}_{b,3}^2$$
$$+ (\alpha + 2\beta \bar{D}_{b,3}) \times N_i D_1 \frac{4\pi r_s^2 (r_3 - r_s)}{V_n'} \quad (12)$$
$$+ N_i \beta D_1^2 \frac{4\pi r_s^3}{V_n'} \frac{(r_3 - r_s)}{r_3}$$

$$\langle N_2(D) \rangle = \alpha \bar{D}_{b,2} + \beta \bar{D}_{b,2}^2$$
$$+ (\alpha + 2\beta \bar{D}_{b,2}) \times N_i D_1 \frac{2\pi r_s^2}{A_n'} \ln\left(\frac{r_2}{r_s}\right) \quad (13)$$
$$+ N_i \beta D_1^2 \frac{\pi r_s^2}{A_n'}$$

Here it was taken into account, that the volumes or areas of the GNPs must be excluded when calculating the integrals in Eqs. (2) and (3). The corrected values of nucleus volume $V_n'$ and area $A_n'$ are given in Eq. (14).

$$V_n' = \frac{4\pi}{3}(R_n^3 - N r_s^3) \quad ; \quad A_n' = \pi(R_n^2 - N r_s^2) \quad (14)$$

It is important to note that also the contributions from the background dose are different between the 3D and 2D cases. Generally, $\bar{D}_{b,2}$ is larger than $\bar{D}_{b,3}$ if the number of NPs undergoing an interaction is higher than 1 because then $r_2$ is smaller than $r_3$ (cf. Eq. (S1.16) in Supplement 1).

*Numerical example*

It may be useful to consider concrete values by using the parameters from the work of Velten et al., namely $\alpha = 0.019$ Gy$^{-1}$, $\beta = 0.052$ Gy$^{-2}$, $r_s = 25$ nm, and $R_n = 4$ μm.

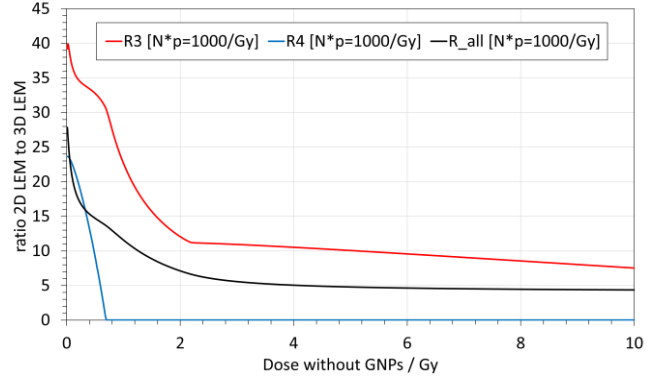

Fig. 1: Ratios $R_3$ and $R_4$ of the third and fourth terms in Eq. (13) to the third and fourth terms in Eq. (12) as a function of dose without GNPs (red and blue curve) for the model parameters considered in the text. The black curve shows the ratio of the total number of lesions predicted by Eq. (13) to the value obtained with Eq. (12).

Assume that 1000 GNPs undergo an ionizing interaction. (This corresponds approximately to the case of 80 keV photons at a dose without GNPs of $D_0 = 1$, cf. Table 1 in [1]). With an assumed surface dose of the GNP of $D_1 = 100$ Gy, $\bar{D}_{b,3}$ and $\bar{D}_{b,2}$ are about 8.6 Gy and 26.5 Gy, respectively. The resulting predicted number of lesions is 5.30 and 61.3, respectively. Here, it was taken into account that the background dose in the 2D case exceeds the threshold dose of 20 Gy for the transition from the linear-quadratic to a linear dose dependence. (See Eq. (S2.9) in Supplement 2 for the expression replacing Eq. (13) in this case.) That is, the 2D approach predicts more than 10 times more lesions than the 3D approach in this case.

The contribution of the background dose to the predicted lesions is 4.02 in the 3D LEM, while 1.17 and 0.11 lesions are coming from the third and fourth terms in Eq. (12), respectively. In the LEM in 2D case, the background dose accounts for 34.7 lesions, and the dose peaks around GNPs contribute 26.6 lesions. These numbers corroborate that a significant overestimation of the number of lesions occurs in the "LEM in 2D" when the 3D dose kernels are used. Of course, the concrete numbers are only valid for the specific case and the assumptions made in the derivation, such as the $1/r^2$ dose dependence.

The number of interacting GNPs increases proportional to the incident fluence (or dose without GNPs). At higher numbers of interacting GNPs, the contribution of the background dose grows faster than the LEM-specific terms in Eqs. (12) and (13) relating to dose non-uniformity (cf. Supplement 3). However, the overall overestimation of the number of lesions remains, as can be seen in Fig. 1.

The red curve in Fig. 1 relates to the LEM-specific third term of Eqs. (12) and (13). Here, the ratio between the 2D version and the 3D version of the LEM is over a factor of 10 over most of the range of dose without GNP. (The ratio of the fourth terms rapidly drops to zero when the background dose of the 2D case exceeds the threshold dose for transition from the linear-quadratic to a linear dose dependence.) The ratio of the total number of lesions from the 2D and 3D models settles around 5 in Fig. 1.

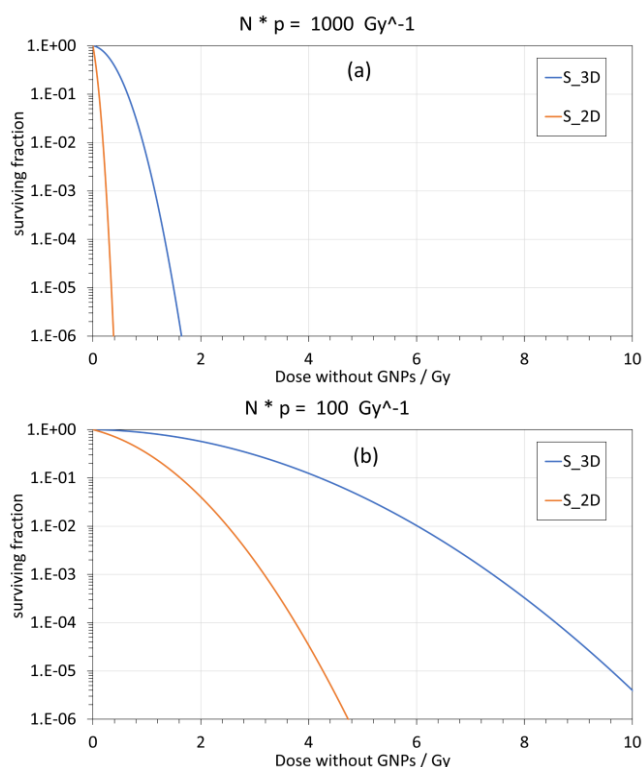

Fig. 2: Survival curves obtained with the 3D and 2D versions of the LEM for the model parameters used by Velten et al. [1] and for (a) 1000 and (b) 100 GNPs undergoing an ionization at a dose without GNPs of 1 Gy.

How this overestimation of the number of lesions affects the predicted survival curves can be seen in Fig. 2, where the upper panel refers to the case considered so far, namely that 1000 GNPs undergo an ionizing interaction at a dose of 1 Gy. The bottom panel refers to the case of only 100 ionizations in GNPs at 1 Gy. This could occur due to different energies of the incident radiation or when lower concentrations of GNPs are considered. In both cases, it can be seen in Fig. 2 that the LEM in 2D predicts a much lower surviving fraction than the 3D version of the LEM.

### Concerns regarding "macroscopic dose" estimates

Absorbed dose is a point quantity. Therefore, it is reasonable to assume that what Velten et al. and Lin et al. [2] refer to as the "macroscopic dose" is the macroscopic average dose. More precisely, it appears to be the macroscopic average dose to water in the absence of GNPs.

It is not clear from either work whether an isolated cell loaded with GNPs and surrounded by cells without GNPs is being considered or a tumor loaded with GNPs. In the latter case, all but the cells at the tumor's surface would be in an environment with the same gold concentration as the cell under consideration itself. In a more realistic scenario, where the tissue adjacent to the tumor also contains GNPs, the preceding statement applies to all tumor cells.

The macroscopic average dose in a piece of tissue with a given concentration of gold in the form of GNPs is enhanced by the increased absorption of gold compared to the elements that tissue is composed of. For the photon spectra shown by Velten et al., the macroscopic average dose can be estimated from the mass concentration of gold and from literature data on mass energy-absorption coefficients [4].

For the photon irradiations, $1.4 \times 10^5$ GNPs of 50 nm diameter were considered, which have a total volume of about 9 μm³. This corresponds to 3.4 % of the volume of the nucleus (8 μm diameter sphere) and 0.7 % of the volume of the cell (13.5 μm diameter). The corresponding mass fractions of gold are 660 g/kg and 137 g/kg.

The gold-to-water ratio of the mass energy-absorption coefficients for the monoenergetic photon energies ranges from 11 (250 keV) to 145 (50 keV). If the weights of the different photon energies in the mock 250 kVp spectra are used (given in Table S1 of [1]), this ratio is between 58 and 87. Therefore, the macroscopic average dose under secondary particle equilibrium for the case of the GNPs uniformly distributed over the cells is expected to be higher. This dose is expected to be higher by a factor between 2.4 (250 keV) and 20.7 (50 keV) – or between 8.8 and 12.8 for the mock 250 kVp spectra – than for the case without GNPs.

These factors are even higher than the radiation enhancement factors (ratio between the doses leading to the same survival rate) seen in the left part of Fig. 5 in [1]. The reason for this is as follows. In the approach of Velten et al. the proportion of the energy of electrons produced in GNPs, which is imparted outside the cell, is not compensated by energy imparted in a realistic irradiation scenario by electrons from radiation interactions in GNPs located in neighboring cells.

In the other two cases, where the GNPs are located only in the cell nucleus and only in the cytoplasm, the dose inside the nucleus differs from that in the cytoplasm. In fact, there is a dose gradient at the interface between the two regions. In the simple geometry of concentric spheres, the variation of the absorbed dose with radial distance from the center of the spheres could be calculated from the data shown in Fig. 1 of [1].

In any case, the "macroscopic dose" plotted on the $x$-axes in Fig. 5 and Fig. S6 of [1] is not the macroscopically averaged dose for the curves with GNPs and should be called differently. For instance, the incident photon fluence could be used instead. Or it should be clearly stated that it is the dose in the absence of GNPs as has been done with Fig. 1 and Fig. 2 of this paper. (For comparing the outcomes of the two models, the use of dose without GNPs as independent variable was a necessity.)

### Conclusions

It is understood that Velten et al. [1] intended to reproduce the work of Lin et al. [2] and add some new insights into the contributions of photons slightly above the gold K-edge. The author fully endorses their recommendations to report more details on the photon spectra used in publications. Such details are a prerequisite for any attempt to check whether other codes give the same results.

Insufficient details on the methods used make it difficult for other researchers to assess or verify the results.

Unfortunately, like Lin et al. before them, Velten et al. also fail to provide sufficient details on their modifications to the LEM or to justify why this differing approach should give correct predictions. From the analysis given in this Comment, it appears that the "LEM approach in 2D" as used by Velten et al. can lead to survival rates that are underestimated by orders of magnitude.

Another recommendation made here is to only use well-defined terms or to clearly state what is meant, for instance, by "macroscopic dose". Plotting results for cell survival in the presence of GNPs as a function of the macroscopic average dose in their absence is justified for low gold concentrations. For gold concentrations as high as 660 g/kg and 137 g/kg (which appear unrealistic compared to values reported in radiobiological studies [5]), this approach is misleading. The use of the average dose with GNPs seems better suited to disentangle the effects of dose enhancement and "radio-enhancement" [6], that is, the increase in average dose versus local dose (that affects cell survival predicted by the LEM).

## Acknowledgements

The author acknowledges the feedback received from the two unknown colleagues who reviewed the paper when it was submitted to Biomedical Physics and Engineering Express. One of them prompted the more detailed discussion on what exactly is the problem when using the 2D LEM with a cut through the three-dimensional dose distribution instead of a transformed distribution. The second reviewer noted typos in the formulas and stimulated the extension of the paper to also include estimates for the case of many GNPs.

# Comment on "Reproducibility study of Monte Carlo simulations for nanoparticle dose enhancement and biological modeling of cell survival curves" by Velten et al. [Biomed Phys Eng Express 2023;9:045004]

# Supplement 1

**Hans Rabus**
Physikalisch-Technische Bundesanstalt, Berlin, Germany
**Email**: hans.rabus@ptb.de

In this Supplement, the mathematical expressions for the average dose in the cell nucleus and the background dose in the vicinity of a nanoparticle (NP) are derived for the case that

- the NPs are uniformly distributed inside the nucleus,
- the extra dose from an interaction in the NP varies with the inverse square of the radial distance from the center of the NP.

The nucleus is assumed to be a sphere and a circle in the 3D model and 2D model, respectively. Both sphere and circle have a radius of $R_n$. Then the number densities of NPs undergoing an ionizing interaction in the two cases, $n_3$ and $n_2$, are given by Eq. (S1.1).

$$n_3 = \frac{3NpD_0}{4\pi R_n^3} \quad ; \quad n_2 = \frac{NpD_0}{R_n^2 \pi} \tag{S1.1}$$

Here, $N$ is the total number of NPs in the nucleus, $D_0$ is the dose in the absence of NPs, and $p$ is the probability of a NP undergoing a photon interaction for a photon fluence producing a dose of 1 Gy in the absence of NPs. ($p$ corresponds to the parameter $P_{tot}$ in the work of Velten et al. [1].)

The additional dose $D_i$ from a single NP undergoing an ionizing interaction is given by Eq. (S1.2).

$$D_i(r) = D_1 \frac{r_s^2}{r^2} \tag{S1.2}$$

Here $D_1$ is the dose contribution at the surface of the NP due to an ionization in the NP, and $r_s$ is the radius of the spherical NP. The average doses at a point in the nucleus are given by Eqs. (S1.3) and (S1.4) for the 3D case and 2D case, respectively, where $D_0$ is the dose in the absence of NPs.

$$D_{a,3} = D_0 + \iint D_1 \frac{r_s^2}{r^2} n_3 r^2 dr\, d\Omega = D_0 + D_1 r_s^2 n_3 \int \left( \int dr \right) d\Omega \tag{S1.3}$$

$$D_{a,2} = D_0 + \iint D_1 \frac{r_s^2}{r^2} n_2 r\, dr\, d\varphi = D_0 + D_1 r_s^2 n_2 \int \left( \int \frac{1}{r} dr \right) d\varphi \tag{S1.4}$$

The support of the inner integrals appearing on the right-hand sides of Eqs. (S1.3) and and (S1.4) is between $r_s$ and the length $l$ of the chord from the point where the dose is evaluated to the surface of the nucleus along the direction specified by the polar angle $\theta$ and plane angle $\varphi$ in the 3D case and 2D case, respectively.

$$D_{a,3} = D_0 + D_1 r_s^2 n_3 \int_0^\pi [l(\theta) - r_s] 2\pi \sin\theta\, d\theta = D_0 + NpD_0 D_1 \frac{r_s^2}{R_n^2} \times \frac{3}{R_n} \left( \frac{1}{2} \int_0^\pi l(\theta) \sin\theta\, d\theta - r_s \right) \tag{S1.5}$$

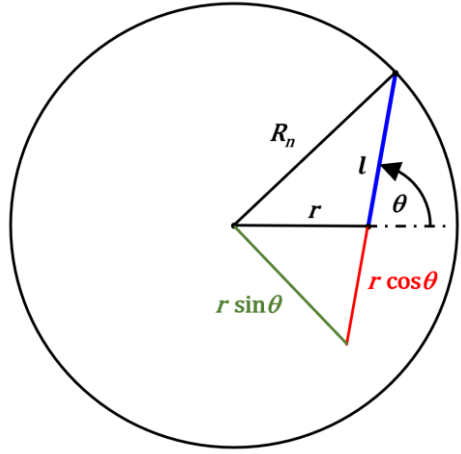

Fig. S1.1: Illustration of the determination of the mean chord from a point to the surface of a sphere (or to the perimeter of a circle in the 2D case).

$$D_{a,2} = D_0 + D_1 r_s^2 n_2 \int_0^{2\pi} \ln \frac{l(\varphi)}{r_s} d\varphi = D_0 + Np D_0 D_1 \frac{r_s^2}{R_n^2} \times \frac{1}{\pi} \int_0^{2\pi} \ln \frac{l(\varphi)}{r_s} d\varphi \tag{S1.6}$$

The functional form of the expression for the length of the chord is given by Eq. (S1.7) (Fig. S1.1; in the 2D case, $\theta$ must be replaced by $\varphi$).

$$l(\theta, r) = -r \cos \theta + \sqrt{R_n^2 - r^2 (\sin \theta)^2} \tag{S1.7}$$

Here, $r$ is radial distance of the point at which the dose is evaluated from the center of the nucleus. In the 3D case, the integral can be calculated analytically, and the last factor in Eq. (S1.5) is obtained as the expression given in Eq. (S1.8).

$$f_3(r) = \frac{3}{R_n} \left( \frac{1}{2} \int_0^\pi l(\theta) \sin \theta \, d\theta - r_s \right) = 3 \left( \frac{1}{2} + \frac{1}{4} \left( \frac{R_n}{r} - \frac{r}{R_n} \right) \ln \left( \frac{R_n + r}{R_n - r} \right) - \frac{r_s}{R_n} \right) \tag{S1.8}$$

The mean value of $f_3$ is given by

$$\bar{f}_3 = \frac{3}{R_n^3} \int_0^{R_n} f_3(r) r^2 dr = \frac{3}{R_n^3} \int_0^{R_n} 3 \left( \frac{1}{2} + \frac{1}{4} \left( \frac{R_n}{r} - \frac{r}{R_n} \right) \ln \left( \frac{R_n + r}{R_n - r} \right) - \frac{r_s}{R_n} \right) r^2 dr$$
$$= 3 \left( \frac{1}{2} - \frac{r_s}{R_n} \right) + \frac{9}{4} \int_0^1 (x - x^3) \ln \left( \frac{1 + x}{1 - x} \right) dx \tag{S1.9}$$

Where in the last integral $r_s/R_n$ has been substituted with $x$. This integral can be calculated analytically and gives a value of 1/3 (see Appendix, page S1-5), so that finally the mean dose in the 3D case is obtained by Eq. (S1.10).

$$\bar{D}_{a,3} = D_0 + Np\, D_0 D_1 \frac{r_s^2}{R_n^2} \times \left( \frac{9}{4} - 3 \frac{r_s}{R_n} \right) \tag{S1.10}$$

The corresponding factor for the 2D case

$$f_2(r) = \frac{1}{\pi} \int_0^{2\pi} \ln \frac{l(\varphi)}{r_s} d\varphi = \frac{1}{\pi} \int_0^{2\pi} \ln \frac{l(\varphi)}{R_n} d\varphi + 2 \ln \frac{R_n}{r_s} \tag{S1.11}$$

cannot be obtained analytically. The respective data shown as red curve in Fig. 1 were determined by Monte Carlo integration using an ad-hoc GNU data language (GDL) script assuming the parameter values $r_s = 25$ nm and $R_n = 4$ μm. This script was also used for calculating the contribution to $\bar{f}_2$ which is independent of $r_s$. The resulting expressions for $\bar{f}_2$ and $\bar{D}_{b,2}$ are given in Eqs. (S1.12) and (S1.13).

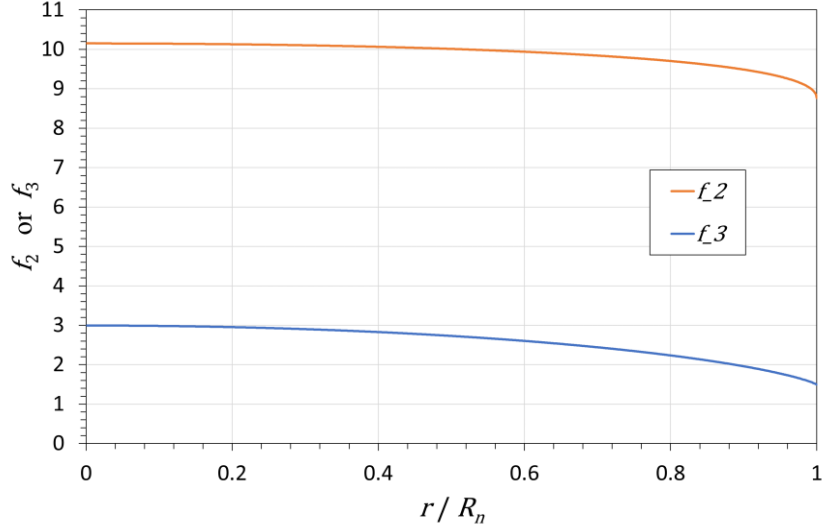

Fig. S1.2: Parameters $f_3$ and $f_2$ as a function of $r/R_n$ for the parameter values $r_s = 25$ nm and $R_n = 4$ μm.

$$\bar{f}_2 = -0.387 + 2\ln\frac{R_n}{r_s} = 2\ln\left(0.824\frac{R_n}{r_s}\right) \tag{S1.12}$$

$$\bar{D}_{a,2} = D_0 + NpD_0D_1\frac{r_s^2}{R_n^2} \times 2\ln\left(0.824\frac{R_n}{r_s}\right) \tag{S1.13}$$

Fig. S1.2 shows the functional dependences of $f_3$ and $f_2$ on $r$ as the blue and red curves, respectively, for the parameter values $r_s = 25$ nm and $R_n = 4$ μm.

The background dose in the vicinity of an NP undergoing an ionizing interaction produced by the other interacting NPs can be obtained in a similar way by considering only NP locations at radial distances exceeding those of the radius of a sphere of volume $V_n/N_i$ (or of a circle of area $A_n/N_i$) centered at the considered NP.

Here

$$N_i = NpD_0 \tag{S1.14}$$

is the number of NPs undergoing an ionizing interaction, and the radius of the sphere, $r_3$, and the radius of the circle, $r_2$, are given by:

$$r_3 = \frac{R_n}{N_i^{1/3}} \quad ; \quad r_2 = \frac{R_n}{N_i^{1/2}} \tag{S1.15}$$

Evidently, $r_2$ is always smaller than $r_3$ when $N_i$ is greater than 1. These radii correspond to half of the nearest-neighbor distance between of NPs undergoing an ionizing interaction and replace $r_s$ in Eqs. (S1.8) and (S1.12). The resulting expressions for the background doses are given in Eqs. (S1.16) and (S1.17).

$$D_{b,3} = D_0 + N_iD_1\frac{r_s^2}{R_n^2} \times \left(f_3(r) + \frac{3r_s}{R_n} - \frac{3}{N_i^{1/3}}\right) \tag{S1.16}$$

$$D_{b,2} = D_0 + N_iD_1\frac{r_s^2}{R_n^2} \times \left(f_2(r) - 2\ln\frac{R_n}{r_s} + \ln(N_i)\right) \tag{S1.17}$$

The mean background doses are given by Eqs. (S1.18) and (S1.19).

$$\bar{D}_{b,3} = D_0 + N_iD_1\frac{r_s^2}{R_n^2} \times \left(\frac{9}{4} - \frac{3}{N_i^{1/3}}\right) \tag{S1.18}$$

$$\bar{D}_{b,2} = D_0 + N_iD_1\frac{r_s^2}{R_n^2} \times \ln(0.68N_i) \tag{S1.19}$$

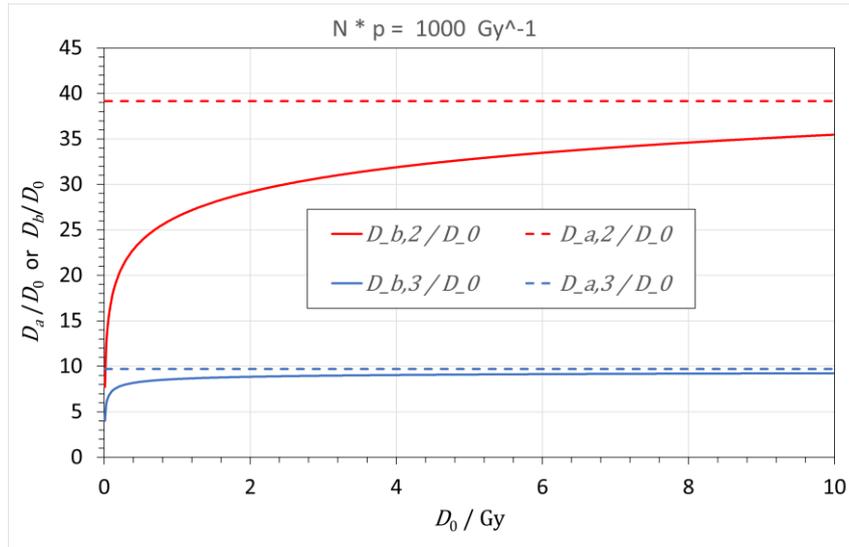

Fig. S1.3: Dependence of the ratios of average doses and background doses to the dose $D_0$ without NPs on $D_0$. The data correspond to the case when the number of NPs undergoing an ionizing interaction at a dose of 1 Gy is 1000. This is corresponds approximately to the case of 80 keV photons in the work of Velten et al. [1].

It should be noted that the average doses given by Eqs. (S1.10) and (S1.13) are strictly proportional to the dose $D_0$ in the absence of NPs, whereas the ratio of the background doses given by Eqs. (S1.18) and (S1.19) to $D_0$ are not constants. This is shown in Fig. S1.3 for the case of 1000 interacting NPs at a dose of 1 Gy.

## Appendix: Evaluation of the last integral in Eq. (S1.9)

The integral can be rewritten as follows:

$$\int_0^1 (x - x^3) \ln\left(\frac{1+x}{1-x}\right) dx = \int_0^1 (x - x^3)[\ln(1+x) - \ln(1-x)] dx$$

$$= \int_0^1 [-(1+x)^3 + 3(1+x)^2 - 2(1+x)] \ln(1+x) \, dx \qquad (S1.20)$$

$$- \int_0^1 [(1-x)^3 - 3(1-x)^2 + 2(1-x)] \ln(1-x) \, dx$$

As can be simply verified by calculating the derivative, the primitives of $y^n \ln y$ are

$$\int y^n \ln y \, dy = \frac{1}{n+1} y^{n+1} \ln y - \frac{1}{(n+1)^2} y^{n+1} \qquad (S1.21)$$

So that

$$\int_0^1 (x - x^3) \ln(1+x) \, dx$$

$$= \left(\left[-\frac{1}{4}(1+x)^4 + (1+x)^3 - (1+x)^2\right] \ln(1+x)\right)\Big|_0^1 \qquad (S1.22)$$

$$- \left(-\frac{1}{16}(1+x)^4 + \frac{1}{3}(1+x)^3 - \frac{1}{2}(1+x)^2\right)\Big|_0^1 = 0 - 0 + \frac{1}{3} + \frac{11}{48}$$

$$\int_0^1 (x - x^3) \ln(1-x) \, dx$$

$$= \left(\left[-\frac{1}{4}(1-x)^4 + (1-x)^3 - (1-x)^2\right] \ln(1-x)\right)\Big|_0^1 \qquad (S1.23)$$

$$- \left(-\frac{1}{16}(1-x)^4 + \frac{1}{3}(1-x)^3 - \frac{1}{2}(1-x)^2\right)\Big|_0^1 = 0 - 0 + 0 + \frac{11}{48}$$

And finally

$$\int_0^1 (x - x^3) \ln\left(\frac{1+x}{1-x}\right) dx = \frac{1}{3} \qquad (S1.24)$$

# Comment on "Reproducibility study of Monte Carlo simulations for nanoparticle dose enhancement and biological modeling of cell survival curves" by Velten et al. [Biomed Phys Eng Express 2023;9:045004]

## Supplement 2

**Hans Rabus**
Physikalisch-Technische Bundesanstalt, Berlin, Germany
**Email**: hans.rabus@ptb.de

In this Supplement, the mathematical expressions for the modifications of Eqs. (12) and (13) in the main document are derived for the case that the dose dependence of the number of induced lesions changes from linear-quadratic to a linear when the dose $D$ exceeds a threshold $D_t$. The derivation refers to spherical nanoparticles (NPs) located inside a spherical nucleus.

The linear dose dependence of the induced lesions at high doses is given by [1]:

$$N(D; \alpha, \beta) = \alpha D_t + \beta D_t^2 + (\alpha + 2\beta D_t)(D - D_t) \tag{S2.1}$$

where $\alpha$ and $\beta$ are the parameters of the linear-quadratic dose dependence at doses below $D_t$. Eq. (S2.1) can be conveniently rewritten as follows:

$$N(D; \alpha, \beta) = (\alpha + 2\beta D_t)D - \beta D_t^2 \tag{S2.2}$$

In the vicinity of an NP, the dose is assumed to have the following dependence on radial distance $r$ from the center of the NP:

$$D = D_b + D_1 \frac{r_s^2}{r^2} \tag{S2.3}$$

where $D_b$ is the background dose due to interactions in water and the other NPs undergoing an ionizing interaction, which was derived in Supplement 1. $D_1$ is the extra dose at the NP surface due to an ionizing interaction in the NP, and $t_s$ is the radius of the NP.

*Case 1: $D_b \geq D_t$*

In case that the background dose $D_b$ exceeds the threshold dose $D_t$, the dose around an NP is always larger than $D_t$ as well. For a dose dependence as given in Eq. (S2.3) the following expression is obtained for the predicted number of lesions in the 3D case:

$$\langle N_3(D) \rangle = N_i \times \frac{1}{V_n - NV_s} \int_{r_s}^{r_3} \left[ (\alpha + 2\beta D_t)\left(D_{b,3} + D_1 \frac{r_s^2}{r^2}\right) - \beta D_t^2 \right] 4\pi r^2 dr \tag{S2.4}$$

Here $N_i$ is the number of NPs undergoing an ionizing interaction, $V_n$ is the volume of the nucleus, $V_s$ is the volume of a NP, and $N$ is the total number of NPs in the nucleus. The upper limit of the integral, $r_3$, is the radius of a sphere having a volume $V_n/N_i$.

The corresponding expression for the 2D case is

$$\langle N_2(D) \rangle = N_i \times \frac{1}{A_n - NA_s} \int_{r_s}^{r_2} \left[ (\alpha + 2\beta D_t)\left(D_{b,2} + D_1 \frac{r_s^2}{r^2}\right) - \beta D_t^2 \right] 2\pi r dr \tag{S2.5}$$

where $A_n$ is the area of the 2D-nucleus, $A_s$ is the cross-sectional area of a NP, and the upper limit of the integral, $r_2$, is the radius of a circle of area $A_n/N_i$.

Evaluating the integrals in Eqs. (S2.4) and (S2.5) gives Eqs. (S2.6) and (S2.7), respectively.

$$\langle N_3(D)\rangle = \frac{N_i}{V_n - NV_s}\left[\left((\alpha + 2\beta D_t)D_{b,3} - \beta D_t^2\right) \times \frac{4\pi}{3}(r_3^3 - r_s^3) + (\alpha + 2\beta D_t) \times D_1 4\pi r_s^2(r_3 - r_s)\right] \quad (S2.6)$$

$$\langle N_2(D)\rangle = \frac{N_i}{A_n - NA_s}\left[\left((\alpha + 2\beta D_t)D_{b,2} - \beta D_t^2\right) \times \pi(r_2^2 - r_s^2) + (\alpha + 2\beta D_t) \times D_1 2\pi r_s^2 \ln\left(\frac{r_2}{r_s}\right)\right] \quad (S2.7)$$

Using the definitions of $r_2$ and $r_3$, Eqs. (S2.8) and (S2.9) are obtained as the final expressions.

$$\langle N_3(D)\rangle = (\alpha + 2\beta D_t)D_{b,3} - \beta D_t^2 + N_i(\alpha + 2\beta D_t)D_1 \frac{3r_s^2(r_3 - r_s)}{R_n^3 - Nr_s^3} \quad (S2.8)$$

$$\langle N_2(D)\rangle = (\alpha + 2\beta D_t)D_{b,2} - \beta D_t^2 + N_i(\alpha + 2\beta D_t)D_1 \frac{2\pi r_s^2}{R_n^2 - Nr_s^2}\ln\left(\frac{r_2}{r_s}\right) \quad (S2.9)$$

*Case 2: $D_b < D_t$*

If the background dose $D_b$ is below the threshold dose $D_t$, then the dose in the nucleus exceeds $D_t$ only within radial distances from a NP of below a threshold value $r_t$ given by Eq. (S2.10).

$$r_t = r_s \times \sqrt{\frac{D_1}{D_t - D_b}} \quad (S2.10)$$

If this threshold value is larger than $r_2$ or $r_3$ in the 2D or 3D model, respectively, then the dose inside the corresponding sphere or circle around an interacting NP always exceeds the threshold dose. Therefore, Eqs. (S2.8) are (S2.9) also describe this case. However, to distinguish between the lesions due to the background dose alone and the contributions due to dose-non-uniformity, it is convenient to rewrite the expressions as in Eqs. (S2.11) and (S2.12).

$$\langle N_3(D)\rangle = \alpha D_{b,3} + \beta D_{b,3}^2 - \beta\left(D_{b,3} - D_t\right)^2 + N_i(\alpha + 2\beta D_t)D_1 \frac{3r_s^2(r_3 - r_s)}{R_n^3 - Nr_s^3} \quad (S2.11)$$

$$\langle N_2(D)\rangle = \alpha D_{b,3} + \beta D_{b,2}^2 - \beta\left(D_{b,2} - D_t\right)^2 + N_i(\alpha + 2\beta D_t)D_1 \frac{2\pi r_s^2}{R_n^2 - Nr_s^2}\ln\left(\frac{r_2}{r_s}\right) \quad (S2.12)$$

For a threshold radius smaller than the radius of the sphere or circle associated with the NP, Eqs. (S2.4) and (S2.5) are replaced by Eqs. (S2.13) and (S2.14), respectively.

$$\langle N_3(D)\rangle = \frac{N_i}{V_n - NV_s}\left[\int_{r_s}^{r_{t,3}}\left[(\alpha + 2\beta D_t)\left(D_{b,3} + D_1\frac{r_s^2}{r^2}\right) - \beta D_t^2\right]4\pi r^2 dr \right.$$
$$\left. + \int_{r_{t3}}^{r_3}\left[\alpha\left(D_{b,3} + D_1\frac{r_s^2}{r^2}\right) + \beta\left(D_{b,3} + D_1\frac{r_s^2}{r^2}\right)^2\right]4\pi r^2 dr\right] \quad (S2.13)$$

$$\langle N_2(D)\rangle = \frac{N_i}{A_n - NA_s}\left[\int_{r_s}^{r_{t,2}}\left[(\alpha + 2\beta D_t)\left(D_{b,2} + D_1\frac{r_s^2}{r^2}\right) - \beta D_t^2\right]2\pi r dr \right.$$
$$\left. + \int_{r_{t,2}}^{r_2}\left[\alpha\left(D_{b,2} + D_1\frac{r_s^2}{r^2}\right) + \beta\left(D_{b,2} + D_1\frac{r_s^2}{r^2}\right)^2\right]2\pi r dr\right] \quad (S2.14)$$

Evaluating the integrals gives Eqs. (S2.15) and (S2.16).

$$\langle N_3(D)\rangle = \frac{N_i}{V_n - NV_s}\left[(\alpha D_{b,3} + \beta D_{b,3}^2) \times \frac{4\pi}{3}(r_3^3 - r_s^3) - \beta(D_{b,3} - D_t)^2 \times \frac{4\pi}{3}(r_{t,3}^3 - r_s^3)\right.$$
$$+ D_1 4\pi r_s^2(\alpha(r_3 - r_s) + 2\beta[D_t(r_{t,3} - r_s) + D_{b,2}(r_3 - r_{t3})])$$
$$\left. + \beta D_1^2 4\pi r_s^4\left(\frac{1}{r_{t,3}} - \frac{1}{r_3}\right)\right] \tag{S2.15}$$

$$\langle N_2(D)\rangle = \frac{N_i}{A_n - NA_s}\left[(\alpha D_{b,3} + \beta D_{b,3}^2) \times \pi(r_2^2 - r_s^2) - \beta(D_{b,2} - D_t)^2 \times \pi(r_{t,2}^2 - r_s^2)\right.$$
$$\left. + D_1 2\pi r_s^2\left(\alpha \ln\left(\frac{r_2}{r_s}\right) + 2\beta\left[D_t \ln\left(\frac{r_{t,2}}{r_s}\right) + D_{b,2} \ln\left(\frac{r_2}{r_{t,2}}\right)\right]\right) + D_1^2 \pi r_s^4\left(\frac{1}{r_{t,2}^2} - \frac{1}{r_2^2}\right)\right] \tag{S2.16}$$

This gives finally gives the second lines in Eqs. (S2.17) and (S2.18).

*Summary*

In summary, the expressions for the number of lesions predicted by the 2D and 3D model are given by Eqs. (S2.17) and (S2.18).

$$\langle N_3(D)\rangle = \begin{cases} (\alpha + 2\beta D_t)D_{b,3} - \beta D_t^2 + N_i(\alpha + 2\beta D_t)D_1 \dfrac{3r_s^2(r_3 - r_s)}{R_n^3 - Nr_s^3} & \bigg| \begin{array}{l} D_{b,3} \geq D_t \\ \vee\, r_{t,3} \geq r_3 \end{array} \\[2em] \alpha D_{b,3} + \beta D_{b,3}^2 - N_i\beta(D_{b,3} - D_t)^2\dfrac{r_{t,3}^3 - r_s^3}{R_n^3 - Nr_s^3} \\ +N_i(\alpha(r_3 - r_s) + 2\beta[D_t(r_{t,3} - r_s) + D_{b,2}(r_3 - r_{t3})])D_1\dfrac{3r_s^2}{R_n^3 - Nr_s^3} & \bigg| \begin{array}{l} D_{b,3} < D_t \\ \wedge\, r_{t,3} < r_3 \end{array} \\ +N_i\beta D_1^2 \dfrac{3r_s^3}{R_n^3 - Nr_s^3}\left(\dfrac{r_s}{r_{t,3}} - \dfrac{r_s}{r_3}\right) \end{cases} \tag{S2.17}$$

$$\langle N_2(D)\rangle = \begin{cases} (\alpha + 2\beta D_t)D_{b,2} - \beta D_t^2 + N_i(\alpha + 2\beta D_t)D_1 \dfrac{2\pi r_s^2}{R_n^2 - Nr_s^2}\ln\left(\dfrac{r_2}{r_s}\right) & \bigg| \begin{array}{l} D_{b,2} \geq D_t \\ \vee\, r_{t,2} \geq r_2 \end{array} \\[2em] \alpha D_{b,2} + \beta D_{b,2}^2 - N_i\beta(D_{b,2} - D_t)^2\dfrac{r_{t,2}^2 - r_s^2}{R_n^2 - Nr_s^2} \\ +N_i\left(\alpha \ln\left(\dfrac{r_2}{r_s}\right) + 2\beta\left[D_t \ln\left(\dfrac{r_{t,2}}{r_s}\right) + D_{b,2}\ln\left(\dfrac{r_2}{r_{t,2}}\right)\right]\right)D_1\dfrac{2r_s^2}{R_n^2 - Nr_s^2} & \bigg| \begin{array}{l} D_{b,2} < D_t \\ \wedge\, r_{t,2} < r_2 \end{array} \\ +N_i\beta D_1^2 \dfrac{r_s^2}{R_n^2 - Nr_s^2}\left(\left(\dfrac{r_s}{r_{t,3}}\right)^2 - \left(\dfrac{r_s}{r_3}\right)^2\right) \end{cases} \tag{S2.18}$$

# Comment on "Reproducibility study of Monte Carlo simulations for nanoparticle dose enhancement and biological modeling of cell survival curves" by Velten et al. [Biomed Phys Eng Express 2023;9:045004]

# Supplement 3

**Hans Rabus**
Physikalisch-Technische Bundesanstalt, Berlin, Germany
**Email**: hans.rabus@ptb.de

In this Supplement, the dependence of the number of lesions predicted by the LEM (Fig. S3.1(a)) and the "LEM in 2D" approach of Velten et al. [1] (Fig. S3.1(b)) on the dose without GNPs is shown for the case of 1000 ionizations in GNPs at a dose of 1 Gy. The data were calculated using Eqs. (S2.17) and (S2.18) in Supplement 2.

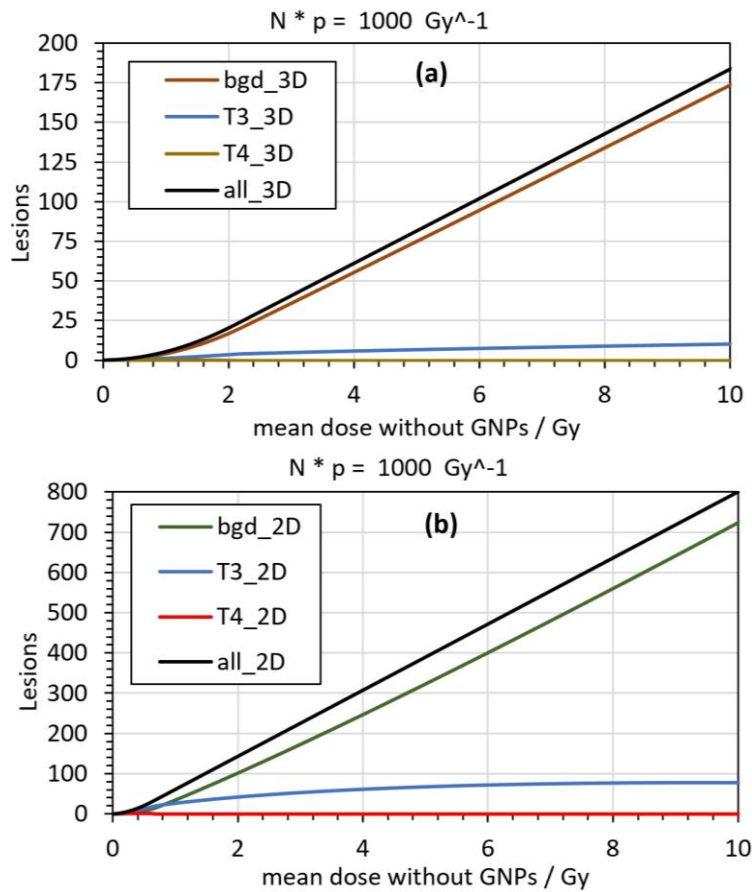

Fig. S3.1: Dependence of the total number of lesions predicted by the 3D LEM and the LEM in 2D and the contirbutions originating in the background dose (bgd) and the linear (T3) and quadratic (T4) terms of the peaked dose distribution around GNPs undergoing an interaction.